\documentstyle[12pt]{article}
\newcommand{\be}{\begin{equation}}
\newcommand{\ee}{\end{equation}}
\newcommand{\bea}{\begin{eqnarray}}
\newcommand{\eea}{\end{eqnarray}}
\newcommand{\nn}{\nonumber \\}
\newcommand{\p}[1]{(\ref{#1})}
\newcommand{\lb}{\label}

\begin{document}
\begin{titlepage}
\begin{flushright}
{}
\end{flushright}
\vskip 1.3truecm

\begin{center}
{\Large\bf The Common Origin of Linear and Nonlinear Chiral
\vspace{0.2cm}

 Multiplets in ${\cal N}{=}4$ Mechanics}
\vspace{1.5cm}

{\large\bf F. Delduc$\,{}^a$, E. Ivanov$\,{}^b$,}\\
\vspace{1cm}

{\it a)Laboratoire de Physique de l'ENS Lyon, CNRS UMR 5672,}\\
{\it   46, all\'ee d'Italie, 69364 Lyon Cedex 07, France}\\
{\tt francois.delduc@ens-lyon.fr}
\vspace{0.3cm}

{\it b)Bogoliubov  Laboratory of Theoretical Physics, JINR,}\\
{\it 141980 Dubna, Moscow region, Russia} \\
{\tt eivanov@theor.jinr.ru}\\

\end{center}
\vspace{0.2cm}
\vskip 0.6truecm  \nopagebreak

\begin{abstract}
\noindent Elaborating on previous work ({\tt hep-th/0605211, 0611247}),
we show how the linear and
nonlinear chiral multiplets of ${\cal N}{=}4$ supersymmetric mechanics with
the off-shell content {\bf (2,4,2)}
can be obtained by gauging three distinct two-parameter isometries of
the ``root'' {\bf (4,4,0)} multiplet actions.  In particular, two different
gauge groups, one abelian and one non-abelian, lead, albeit in a disguised
form in the second case, to the same (unique) nonlinear chiral multiplet.
This provides an evidence that no other
nonlinear chiral ${\cal N}{=}4$ multiplets exist.
General sigma model type actions are discussed, together with the restricted
potential terms coming from the Fayet-Iliopoulos terms associated
with abelian gauge superfields. As in our previous work,
we use the manifestly supersymmetric language of ${\cal N}{=}4, d{=}1$
harmonic superspace. A novel point  is
the necessity to use in parallel the $\lambda$ and $\tau$ gauge frames,
with the ``bridges'' between
these two frames playing a crucial role. It is the ${\cal N}{=}4$
harmonic analyticity which, though being non-manifest
in the $\tau$ frame, gives rise to both linear and nonlinear chirality constraints.

\end{abstract}
\vspace{0.7cm}

\newpage

\end{titlepage}
\section{Introduction}
The one-dimensional supersymmetry and related models of supersymmetric (quantum)
mechanics reveal many specific surprising features and, at the same time, have
a lot of links with higher-dimensional theories of current interest (see e.g.
\cite{Rev,Gr} and refs. therein). This motivates many research groups towards thorough study and
further advancing of this subject (see e.g. \cite{Gr,Top}).

Recently, we argued \cite{1,2} that the plethora of relationships between various $d{=}1$
supermultiplets with the same number of fermionic fields, but different
divisions of the bosonic fields into physical and auxiliary ones
(so called ``$d{=}1$ automorphic dualities'' \cite{ADu}),
can be adequately understood in the approach based on the gauging
of isometries of the invariant actions of some basic (``root'') multiplets by
non-propagating (``topological'') gauge multiplets (these isometries should commute
with supersymmetry, i.e. be triholomorphic). The key merit of our approach
is the possibility to study these relationships in a manifestly supersymmetric
superfield manner, including the choice of supersymmetry-preserving gauges.
Previous analysis \cite{ADu,root} was basically limited to the component level and used
some ``ad-hoc'' substitutions of the auxiliary fields. In the framework of the gauging
procedure, doing this way corresponds to making use of the Wess-Zumino-type gauges.

In \cite{1,2} we focused on the case of ${\cal N}{=}4$ supersymmetric mechanics
and showed that the actions of the ${\cal N}{=}4$ multiplets with the off-shell
contents ${\bf (3,4,1)}$, ${\bf (1, 4,3)}$ and ${\bf (0, 4, 4)}$ can be obtained
by gauging certain isometries of the general actions of the ``root'' multiplet ${\bf (4,4,0)}$ in
the ${\cal N}{=}4, d{=}1$ harmonic superspace. Based on this, we argued that the latter is the
underlying superspace for all ${\cal N}{=}4$ mechanics models. Here we confirm this by
studying, along the same line,
the remaining ${\cal N}{=}4$ multiplets, the chiral and nonlinear chiral off-shell multiplets with
the field content ${\bf (2,4,2)}\,$. They prove to naturally arise as a result of gauging
some two-parameter isometry groups admitting a realization on the harmonic analytic
superfield $q^{+ a}$ which describes
the multiplet ${\bf (4,4,0)}$. The origin of the difference between these two versions of
the ${\bf (2,4,2)}$ multiplet is attributed to the fact that they emerge from  gauging
two essentially different isometries: the linear multiplet is associated with gauging
of some purely shift isometries, while the nonlinear one corresponds to gauging
the product of two ``rotational''
isometries, viz. the target space rescalings and a $U(1)$ subgroup of the $SU(2)$
Pauli-G\"ursey group. We also
recover the same nonlinear version of this multiplet, though in disguise,
by gauging a mixture of
the rescaling and shift isometries. It is the last possible two-parameter
symmetry implementable on $q^{+ a}\,$.
Thus we show that two known off-shell forms of the ${\cal N}{=}4, d{=}1$
multiplet ${\bf (2, 4, 2)}$, the linear and nonlinear ones, in fact exhaust all possibilities.

It should be emphasized that the existence of non-linear cousins of the basic
$d{=}1$ supermultiplets is one of the most amazing features of extended $d{=}1$ supersymmetry.
In superspace they are described by superfields satisfying some nonlinear versions of the
standard constraints (e.g. of chirality constraints). As a result, the realization
of the corresponding off-shell $d{=}1$ supersymmetry on the component fields of
such supermultiplets is intrinsically nonlinear. The list of such multiplets
known to date includes the nonlinear analogs of the ${\cal N}{=}4$ multiplets ${\bf (4,4,0)}$
\cite{Pol,DK,BKS,1,KS}, ${\bf (3,4,1)}$ \cite{IL,IKL}, ${\bf (2, 4, 2)}$ \cite{IKL},
as well as of the ${\cal N}{=}8$ multiplets ${\bf (4,8,4)}$ \cite{N8nl} and ${\bf (2,8,6)}$ \cite{N8nl2}.
As shown in \cite{1}, within the approach based on the gauging procedure the  difference between
linear and nonlinear ${\bf (3, 4, 1)}$ multiplets originates from the fact that
the first multiplet is related to the gauging of the shift or rotational $U(1)$ symmetries,
and the second one
to the gauging of the target space rescalings.
In the case of the ${\bf (2,4,2)}$ multiplets we find an analogous intimate relation
between the type of multiplet and the $q^{+ a}$ two-parameter symmetry group which
has to be gauged to generate it.

\section{Brief preliminaries}
Throughout the paper we use the same ${\cal N}{=}4, d{=}1$ harmonic superspace (HSS) techniques,
conventions and notation as in \cite{IL,1,2}. The ${\cal N}{=}4$ ``root''
multiplet ${\bf (4,4,0)}$ is described by the harmonic analytic superfield $q^{+ a}(\zeta, u)$
which is subjected
to the Grassmann harmonic and bosonic harmonic constraints
\be
\mbox{(a)}\; D^+ q^{+ a} = \bar D^+ q^{+ a} = 0\,, \quad \mbox{(b)} \;D^{++} q^{+ a} = 0\,. \label{qcond}
\ee
Here $(\zeta, u)$ are coordinates of the harmonic analytic ${\cal N}{=}4$ superspace \cite{IL},
$(\zeta, u) = (t_A, \theta^+, \bar\theta^+, u^\pm_i)\,$, $u^{+ i}u_i^- =1\,$, they are related
to the standard ${\cal N}{=}4$ superspace (central basis) coordinates
$z = ( t, \theta_i, \bar\theta^i)$ as
\be
t_A = t -i (\theta^+\bar\theta^- + \theta^-\bar\theta^+), \quad \theta^\pm = \theta^iu^\pm_i\,, \;
 \bar\theta^\pm = \bar\theta^iu^\pm_i\,.
\ee
Respectively, the ${\cal N}{=}4$ covariant spinor derivatives and their harmonic projections
are defined by
\bea
&& D^i = \frac{\partial}{\partial \theta_i} + i\bar\theta^i \partial_t\,, \;\;
\bar D_i = \frac{\partial}{\partial \bar\theta^i} + i\theta_i \partial_t\,, \;\; \overline{(D^i)}
= -\bar D_i\,, \;\;\{D^i, \bar D_k \} = 2i\,\delta^i_k\partial_t\,,
\label{defD2} \\
&& D^\pm = u^\pm_i D^i\,,\quad \bar D^\pm = u^\pm_i \bar D^i\,, \quad
 \; \{D^+, \bar D^- \} = - \{D^-, \bar D^+ \}
= 2i\,\partial_{t_A}\,. \label{defD1}
\eea
In the analytic basis, the derivatives $D^{+}$ and $\bar D^+$ are short,
\be
D^+ = \frac{\partial}{\partial \theta{}^-}\,, \quad \bar D^+ =
-\frac{\partial}{\partial \bar\theta{}^-}\,,
\ee
so the conditions (\ref{qcond}a) become the harmonic Grassmann Cauchy-Riemann conditions stating
that $q^{+ a}$ does not depend on the coordinates $\theta^-, \bar\theta^-$ in this basis.
The analyticity-preserving harmonic derivative $D^{++}$ and its conjugate $D^{--}$
in the analytic basis are given by
\bea
&& D^{++}=\partial^{++}-2i\theta^+\bar\theta^+\partial_{t_{A}}
+\theta^+\frac{\partial}{\partial\theta^-}
+ \bar\theta^+\frac{\partial}{\partial\bar\theta^-}\,, \nn
&& D^{--}=\partial^{--}-2i\theta^-\bar\theta^-\partial_{t_{A}}
+\theta^-\frac{\partial}{\partial\theta^+}
+  \bar\theta^-\frac{\partial}{\partial\bar\theta^+}\,, \quad
\partial^{\pm\pm} = u^{\pm}_i\frac{\partial}{\partial u^{\mp}_i}\,,
\eea
and are reduced to the pure harmonic partial derivatives $\partial^{\pm\pm}$ in the central basis.
They satisfy the commutation relations
\be
[D^{++},D^{--}]= D^{0}\,, \quad [D^0, D^{\pm\pm}] =
\pm 2 D^{\pm\pm}\,, \lb{DharmAl}\\
\ee
where $D^0$ is the operator counting external harmonic $U(1)$ charges. In the analytic basis
it is given by
\be
 D^0 = u^+_{i}\frac{\partial}{\partial u^+_{i}}-u^-_{i}
\frac{\partial}{\partial u^-_{i}}+
\theta^+\frac{\partial}{\partial \theta^+}
+\bar\theta^+\frac{\partial}{\partial \bar\theta^+}
-\theta^-\frac{\partial}{\partial \theta^-}
-\bar\theta^-\frac{\partial}{\partial \bar\theta^-}\,,\label{Dalg}
\ee
while in the central basis it coincides with its pure harmonic part. On the extra doublet index
$a$ of the superfield $q^{+ a}$ the so-called Pauli-G\"ursey group $SU(2)_{PG}$ is realized.
It commutes with the ${\cal N}{=}4$ supersymmetry generators, as distinct from the $R$-symmetry
$SU(2)_R$ group which acts on the doublet indices $i, k$ of the Grassmann and harmonic coordinates,
spinor derivatives and ${\cal N}{=}4$ supercharges.

The free action of $q^{+ a}$ can be written either in the analytic, or the central superspace
\be
S_{q}^{\mbox{\scriptsize free}}
= -\frac{1}{4}\,\int \mu_H\,(q^{+a}q^-_{a})
=\frac{i}{2}\int \mu^{(-2)}_A\,(q^{+a}\partial_t q^+_{a})\,, \lb{Freeq}
\ee
where $q^{- a} \equiv D^{--}q^{+ a}$ and the integration measures are defined as
\begin{eqnarray}
&& \mu_H = dudtd^4\theta=dudt_{A}(D^-\bar D^-)(D^+\bar D^+)=\mu_{A}^{(-2)}(D^+\bar D^+),\nn
&& \mu_{A}^{(-2)}=dud\zeta^{(-2)}
=dudt_{A}d\theta^+d\bar\theta^+=dudt_{A}(D^-\bar D^-)\,. \label{measures}
\end{eqnarray}
The general sigma model-type action of $q^{+ a}$ (with a non-trivial bosonic target space metric)
is given by
\begin{equation}
S_{q}=\int \mu_H \,{\cal L}(q^{+a}, q^{-b}, u^\pm)\,. \lb{qact}
\end{equation}

The constraint \p{qcond} possesses a seven-parameter group of rigid symmetries commuting with
supersymmetry (it includes $SU(2)_{PG}$ as a subgroup) \cite{2}. One can single out the
appropriate subclasses of the general action \p{qact} (including the free action \p{Freeq})
which are invariant with respect to one or another symmetry of this sort. For further
use we give here the full list of non-equivalent two-parameter symmetries.
\vspace{0.3cm}

\noindent{\it Abelian symmetries}
\be
\mbox{(a)}\; \delta q^{+ a} = \lambda_1\, u^{+ a} + \lambda_2\, c^{(ab)} u^{+}_{b}\,; \quad
\mbox{(b)}\; \delta q^{+ a} = \lambda_1\, q^{+ a} - \lambda_2\, c^{(ab)} q^+_b  \,. \lb{Abel}
\ee
\vspace{0.3cm}

\noindent{\it Nonabelian symmetry}
\be
\delta q^{+ a} = \lambda_1\, q^{+ a} + \lambda_2\, u^{+ a}  \,.
\lb{Nonabel}
\ee
\vspace{0.3cm}

\noindent Here the constant triplet $c^{ab}$ is normalized as \footnote{We use the same notation for the
unrelated constant triplets in (\ref{Abel}a) and  (\ref{Abel}b), hoping that this will not give rise
to any confusion.}
\be
c^2 = c^{ab}c_{ab}= 2\,.
\ee
The algebra of the transformations \p{Nonabel} provides an example of two-generator solvable algebra.
All other possible two-parameter symmetry groups listed in \cite{2} can be reduced
to \p{Abel}, \p{Nonabel} by
a redefinition of $q^{+ a}\,$.

In what follows we shall gauge these symmetries and show that this gauging gives rise to
three versions of the ${\cal N}{=}4$ multiplet ${\bf (2, 4, 2)}$, with the corresponding general
actions arising from the appropriate invariant subclasses of the general $q^{+ a}$ action \p{qact}.
It turns out that the standard linear chiral ${\cal N}{=}4$ multiplet emerges as the result of gauging
purely shift isometry (\ref{Abel}a) while the two alternative gaugings give rise to two nonlinear
versions of this multiplet. The nonlinear multiplet obtained from gauging the group (\ref{Abel}b)
is identical to the one discovered in \cite{IKL}. The multiplet obtained from gauging \p{Nonabel},
although looking different, can be identified with the previous one after suitable redefinitions.

\setcounter{equation}{0}
\section{Chiral multiplet}
Our gauging prescriptions are basically the same as in other cases \cite{1,2}.

We start with the gauged version of the transformations (\ref{Abel}a)
\begin{equation}
\delta q^{+a}=\Lambda_{1} \,u^{+a}+\Lambda_{2} c^{ab}\,u^+_{b}\,, \quad c^{ab} = c^{ba}\,,\lb{q+Tran}
\end{equation}
where $\Lambda_{1}$ and $\Lambda_{2}$ are now charge-zero analytic superfields.
The gauge covariantization of the $q^{+a}$ harmonic constraint (\ref{qcond}b) is given by
\begin{equation}
D^{++}q^{+a}-V^{++}_{1}u^{+a}-V^{++}_{2}c^{ab}\,u^+_{b}=0\,, \lb{Covar1}
\end{equation}
where $V^{++}_{1}$ and $V^{++}_{2}$ are analytic gauge superfield transforming as
\begin{equation}
\delta V^{++}_{1}=D^{++}\Lambda_{1}\,,\quad \delta V^{++}_{2}=D^{++}\Lambda_{2}\,. \label{TranV}
\end{equation}

In the cases considered here (as distinct from the cases treated in \cite{1,2}), it is convenient
to make use of the ``bridge'' representation of the gauge superfields \cite{HSS,HSS1}.

The analytic superfields  $V^{++}_{1}$ and $V^{++}_{2}$ may be represented as
\begin{equation}
V^{++}_{1}=D^{++}v_{1}\,,\quad V^{++}_{2}=D^{++}v_{2}\,, \lb{Brid1}
\end{equation}
where $v_1(t, \theta, \bar\theta, u)$ and $v_{2}(t, \theta, \bar\theta, u)$ are non-analytic
harmonic superfields which may be interpreted as ``bridges'' between the analytic and central
superspace gauge groups
(called the $\lambda$ and $\tau$ gauge groups, see \cite{HSS1}). They transform under local shifts as
\begin{equation}
\delta v_{1}=\Lambda_{1}(\zeta,u)-\tau_{1}(t,\theta,\bar\theta)\,,\quad \delta v_{2}
=\Lambda_{2}(\zeta,u)-\tau_{2}(t,\theta, \bar\theta)\,, \label{TranBrid}
\end{equation}
where $\tau_{1}(t,\theta, \bar\theta)$ and $\tau_{2}(t,\theta, \bar\theta)$ are non-analytic gauge
superparameters bearing no dependence on the harmonic variables
\begin{equation}
D^{++}\tau_{1}(t,\theta, \bar\theta)=0\,,\quad D^{++}\tau_{2}(t,\theta, \bar\theta)=0\,.
\end{equation}

We now define the non-analytic doublet superfield $Q^{+a}$ by
\begin{equation}
Q^{+a}=q^{+a}-v_{1}u^{+a}-v_{2}c^{ab}\,u^+_{b}\,.\lb{Q+def}
\end{equation}
As a consequence of \p{Covar1}, \p{Brid1}, it satisfies the simple harmonic constraint
\begin{equation}
D^{++}Q^{+a}=0\,,\label{constf}
\end{equation}
which implies
\begin{equation}
Q^{+a}(t,\theta,\bar\theta, u)=Q^{ba}(t,\theta, \bar\theta)u^+_{b},\quad
\overline{Q^{ba}}=Q_{ba}\,, \lb{DefQ}
\end{equation}
where the superfields $Q^{ba}(t,\theta, \bar\theta)$ are independent of the harmonic variables
and form a real quartet. We also have
\be
Q^{- a} \equiv D^{--} Q^{+ a} = Q^{ba}(t,\theta, \bar\theta)u^-_{b} = q^{- a} - v_1 u^{- a} -
v_2 c^{ab}u^-_b\,, \lb{Q-}
\ee
where
\be
q^{- a} \equiv D^{--}q^{+ a} - V^{--}_1 u^{- a} - V^{--}_2 c^{ab}u^-_b \lb{q-}
\ee
and
\be
V^{--}_1 = D^{--} v_1\,, \; \; V^{--}_2 = D^{--} v_2\,, \quad \delta V^{--}_1 = D^{--}\Lambda_1\,,
\;\; \delta V^{--}_2 = D^{--}\Lambda_2\,. \lb{V--}
\ee
{}From \p{Q-} and \p{V--} one can find the gauge transformation law of the non-analytic harmonic
superfield $q^{- a}$:
\be
\delta q^{-a}=\Lambda_{1} \,u^{-a}+\Lambda_{2} c^{ab}\,u^-_{b}\,. \lb{q-Tran}
\ee
Now it is easy to determine how the superfields  $Q^{\pm a}$ introduced in \p{Q+def}, \p{Q-}
transform under the local shift symmetries. They are inert under the $\lambda$ gauge
transformations and have the following
$\tau $ gauge transformation law
\begin{equation}
\delta Q^{\pm a} = \tau_{1} u^{\pm a}+\tau_{2} c^{ab}\,u^\pm_{b} \; \Leftrightarrow \;
\delta Q^{b a} = \tau_2 c^{ab} - \tau_1 \epsilon^{ba}\,.
\label{trans3}
\end{equation}

As a consequence of the Grassmann analyticity constraints \p{qcond}, the
superfield $Q^{+ a}$  satisfies the following fermionic constraints
\begin{eqnarray}
&& D^+q^{+a}=0\,\,\Leftrightarrow\; D^+Q^{+a}+(D^+v_{1})u^{+a}+(D^+v_{2})c^{ab}\,u^+_{b}=0, \nn
&& \bar D^+q^{+a}=0\,\,\Leftrightarrow \bar D^+Q^{+a}+(\bar D^+v_{1})u^{+a}
+(\bar D^+v_{2})c^{ab}\,u^+_{b}=0\,. \label{cons5}
\end{eqnarray}
It is important that, due to the analyticity of the gauge superfields $V^{++}_{I}$, $I=1,2$,
the fermionic connections $D^+v_{I}$, $\bar D^+v_{I}$ depend linearly on the harmonic variables.
We shall use the notations
\begin{equation}
D^+v_{I}(t,\theta,u)=A^{a}_{I}(t,\theta, \bar\theta)u^+_{a},\quad \bar D^+v_{I}(t,\theta,u)=
- \bar A^{a}_{I}(t,\theta, \bar\theta)u^+_{a}. \label{defA}
\end{equation}
Using \p{DefQ} and \p{defA}, one can rewrite \p{cons5} in the following equivalent form with
no harmonic dependence at all:
\begin{equation}
D^{(a}Q^{b)c}-A_{1}^{(a}\epsilon^{b)c}+A_{2}^{(a}c^{b)c}=0\,,\quad
\bar D^{(a}Q^{b)c}+\bar A_{1}^{(a}\epsilon^{b)c}-\bar A_{2}^{(a}c^{b)c}=0\,. \label{cons7}
\end{equation}

Now let us more closely inspect the transformation laws (\ref{trans3}).
We start by choosing a frame in which the matrix $c^{ab}$ has only one non-vanishing component:
\be
c^{12}=c^{21}=i\,, \quad c^{11}=c^{22}=0\,. \lb{frame}
\ee
In this frame the transformations \p{trans3} look as
\begin{equation}
\delta Q^{\pm 1}= (\tau_{1}-i\tau_{2})\,u^{\pm 1}\,,\quad
\delta Q^{\pm 2} = (\tau_{1}+i\tau_{2})\,u^{\pm 2}\,,
\label{trans1}
\end{equation}
or, in terms of the ${\cal N}{=}4$ superfields
$Q^{ab}(t,\theta, \bar\theta)$ defined in \p{DefQ},
\begin{equation}
\delta Q^{12}=(\tau_{1}+ i\tau_{2})\,,\quad \delta Q^{21}=-(\tau_{1}-i\tau_{2})\,,\quad
\delta Q^{11}=\delta Q^{22}=0\,.
\end{equation}
It is then convenient to choose the unitary-type gauge
\be
Q^{12}=Q^{21}=0\,. \lb{GauGe}
\ee
Now the constraints (\ref{cons7}) determine the spinor connection superfields $A^{a}_{I}$ and their
conjugate $\bar A^{a}_{I}$ in terms of the remaining superfield $Q^{11}\equiv \Phi$ and
its complex conjugate $Q^{22} = \bar\Phi$
\be
A^1_1 = \frac{1}{2}\,D^2 \Phi\,, \; A^2_1 = -\frac{1}{2}\,D^1 \bar\Phi\,, \quad
A^1_2 = \frac{i}{2}\,D^2 \Phi\,, \; A^2_2 = \frac{i}{2}\,D^1 \bar\Phi\,, \quad (\mbox{and c.c.})\,,
\ee
simultaneously with imposing the constraints on these superfields
\begin{equation}
D^1 \Phi=\bar D^1 \Phi=0\,,\quad  D^2\bar\Phi=\bar D^2\bar\Phi=0\,.\label{ChirTw}
\end{equation}
These constraints may be interpreted as twisted chirality conditions. They can be given the standard form
of the chirality conditions by relabelling the spinor derivative $D^i, \bar D_i$ in such
a way that the $R$-symmetry $SU(2)$ acting on the indices $i$ gets hidden, while another
$SU(2)\,$ (which rotates $D^i$ through $\bar D^i\,$), gets manifest \footnote{Both these
$R$-symmetry $SU(2)$ are manifest in the quartet notation
$D^{i\underline{i}} = (D^i, \bar D^i) = (\hat{\bar D}{}^{\underline{i}},
\hat{D}{}^{\underline{i}})\,$,  $\overline{D^{i\underline{i}}} = -D_{i\underline{i}}
= (-\bar D_i, D_i) =
(-\hat{D}_{\underline{i}}, \bar D_{\underline{i}})\,$.}. Thus we have succeeded in deriving
the linear chiral ${\cal N}{=}4, d{=}1$ multiplet ${\bf (2, 4, 2)}$ from the analytic multiplet
${\bf (4, 4,0)}$ by gauging two independent shift isometries realized on the latter.

Let us now examine this correspondence on the level of the invariant actions.
We start with the gauge covariantization of the free action \p{Freeq} of the analytic superfield
$q^{+a}$. In the full superspace the covariantized action reads
\begin{equation}
S_{cov}^{free}=\int \mu_H\,  [\,q^{+a}D^{--}q^+_{a}-2V_{1}^{--}q^{+a}u^+_{a}
-2V_{2}^{--}q^{+a}c_{a}^{b}u^+_{b}
+2(V_{1}^{++}V_{2}^{--}-V_{2}^{++}V_{1}^{--})c^{+-}\,]\,. \lb{FreeqCov}
\end{equation}
It is gauge invariant up to a total harmonic derivative in the integrand,
i.e. it is of the Chern-Simons
type. It may be equivalently rewritten in terms of the superfield $Q^{+a}$
and the bridges $v_{1}$, $v_{2}$
\begin{equation}
S_{cov}^{free}=\int \mu_H\, [\,Q^{+a}D^{--}Q^+_{a}+2v_{1}Q^{+a}u^-_{a}
+2v_{2}Q^{+a}c_{a}^{b}u^-_{b}+(v_{1}V_{2}^{++}
-v_{2}V_{1}^{++})c^{--}+v_{1}^2+v_{2}^2\,]\,.
\label{rontudju}
\end{equation}
In this form it is invariant under both the $\lambda$ and $\tau$ gauge transformations.
In fact, the action can be written as a sum of two terms, the first of which transforms only under
the $\tau$ gauge group, and the second only under the $\lambda$ group
\begin{eqnarray}
&& S_{cov}^{free}=S_{\tau}+S_{\lambda}\,, \nn
&& S_{\tau}=\int \mu_H  \left[ Q^{+a}D^{--}Q^+_{a}-\frac{1}{4}(Q^{+a}u^-_{a}-Q^{-a}u^+_{a})^2
-\frac{1}{4}(Q^{+a}{c_{a}}^{b}u^-_{b}-Q^{-a}{c_{a}}^{b}u^+_{b})^2\right], \label{rontudju1}\\
&& S_{\lambda}=\int \mu_H \left[\frac{1}{4}(q^{+a}u^-_{a}- q^{-a}u^+_{a})^{2}+
\frac{1}{4}(q^{+a}{c_{a}}^{b}u^-_{b}- q^{-a}{c_{a}}^{b}u^+_{b})^{2}
\right. \nonumber\\
&&\left.
\qquad \qquad \quad-\, V_{2}^{++}\,q^{-a}{c_{a}}^{b}u^-_{b}-V_{1}^{++} \, q^{-a}u^-_{a}\right].
\end{eqnarray}
To check the invariance of $S_{\tau}$ we use the  following transformation laws :
\begin{eqnarray}
&& \delta(Q^{+a}u^-_{a}-Q^{-a}u^+_{a})= 2\tau_{1}\,,\quad
\delta(Q^{+a}{c_{a}}^{b}u^-_{b}-Q^{-a}{c_{a}}^{b}u^+_{b})= 2\tau_{2}\,,\nn
&& \delta(Q^{+a}Q^-_{a})= \tau_{1}(Q^{+a}u^-_{a}-Q^{-a}u^+_{a})
+ \tau_{2}(Q^{+a}{c_{a}}^{b}u^-_{b}-Q^{-a}{c_{a}}^{b}u^+_{b})\,.
\end{eqnarray}
It is worthwhile to note that all three terms in the action $S_{\tau}$ are $SU(2)$ singlets
(independent of harmonic variables), so that the harmonic integral $\int du$ is in fact not necessary.
After some simple algebra, making use of the integration by parts with respect
to the harmonic derivatives,
the constraint \p{Covar1} and the definitions \p{Brid1}, \p{q-} and \p{V--},
one can show that, up to a total harmonic derivative,
\be
S_\lambda = \frac{1}{2}\,S^{free}_{cov}\,,
\ee
whence one obtains the representation of the gauge-covariantized $q^+$ action \p{FreeqCov}
solely in terms of the superfields $Q^{\pm a}$:
\be
S^{free}_{cov} = 2 S_\tau = 2 \int \mu_H  \left[ Q^{+a}Q^-_{a}
-\frac{1}{4}(Q^{+a}u^-_{a}-Q^{-a}u^+_{a})^2-\frac{1}{4}(Q^{+a}{c_{a}}^{b}u^-_{b}
-Q^{-a}{c_{a}}^{b}u^+_{b})^2\right]. \lb{ActFreetau}
\ee
Now, the gauge condition \p{GauGe} simply amounts to
\begin{equation}
Q^{+a}u^-_{a}-Q^{-a}u^+_{a}=0\,,\quad Q^{+a}{c_{a}}^{b}u^-_{b}-Q^{-a}{c_{a}}^{b}u^+_{b}=0\,,
\end{equation}
and in this gauge the action \p{ActFreetau} takes the standard form of the free action of the
(twisted) chiral ${\bf (2,4,2)}$ multiplet
\be
S^{free}_{cov} = 2 \int \mu_H \,Q^{+a}Q^-_{a} = 2 \int dt d^4\theta\, \Phi \bar\Phi\,.
\ee

Thus the free action of the chiral ${\cal N}{=}4$ multiplet arises as a particular gauge of the
properly gauge-covariantized free action of the analytic multiplet $q^{+ a}\,$. Note that
this equivalence, like in other cases \cite{1,2}, was shown here in a manifestly ${\cal N}{=}4$
supersymmetric superfield approach, without any need to pass to the components.
It is interesting to note that there exist two more equivalent useful forms of
the action \p{FreeqCov} in terms of the original superfields $q^{\pm a}$:
\be
S^{free}_{cov} =  \int \mu_H  \left[q^{+ a}q^-_a - V^{++}_1 (q^{-a} u^-_a)
- V^{++}_2 (q^{- a}c_a^{\;b}u^-_b) \right] =
\int \mu_H  \,(D^{++} q^{- a})q^-_a\,. \lb{NewFree}
\ee
Checking the gauge invariance of the action in the second form is especially simple:
one uses the transformation law \p{q-Tran} and the fact that $D^{++} q^{-a}$ is analytic in virtue
of the relation
\be
D^{++} q^{- a} = q^{+ a} + V^{++}_1 u^{- a} + V^{++}_2 c^{ab}u^-_b\,.
\ee

Let us now comment on the general sigma-model type action. The only invariant of
the $\lambda$ gauge transformations \p{q+Tran}, \p{q-Tran}  which one can construct
from $q^{+ a}$ and $q^{-a }$ is the quantity $X$ defined as follows
\bea
&& X = (q^{+ a}u^+_a) c^{--} - (q^{- a}u^-_a) c^{++}\,, \; D^{\pm\pm} X =
\pm 2(q^{\pm a}u^\pm_{a}) c^{+-} \mp (q^{+ a}u^-_a + q^{-a}u^+_a)c^{\pm\pm} \,, \nonumber \\
&& D^{++}D^{--} X = 2 X\,, \quad (D^{++})^2 X = (D^{--})^2 X = 0\,,\lb{DefX}
\eea
where
$$
c^{\pm \pm} = c^{ab}u^\pm_a u^\pm_b\,, \quad \mbox{etc}\,.
$$
It admits the equivalent representation, in which its $\tau$ gauge invariance becomes manifest
\be
X =  (Q^{+ a}u^+_a) c^{--} - (Q^{- a}u^-_a) c^{++} = 2 Q^{(ad)}c_d^{\;\;b} u^{+}_{(a}u^-_{b)}
\ee
(the remaining relations in \p{DefX} preserve their form modulo the replacements $q^{\pm a}
\Rightarrow Q^{\pm a}$).  The subclass of the general sigma-model type $q^{+}$ action \p{qact}
invariant under the gauge transformations \p{q+Tran}, \p{q-Tran} is then defined as follows
\be
S_{cov} =  \int \mu_H\,  {\cal L}(X, D^{++}X, D^{--}X, u^\pm)\,. \lb{CovGen}
\ee
In the gauge \p{GauGe} and in the frame \p{frame} we have
\be
X = 2i \left(\Phi u^+_1u^-_1 - \bar\Phi u^+_2u^-_2\right)
\ee
and, after integration over harmonics, the action \p{CovGen} is reduced to the most general
action of the twisted chiral superfields $\Phi, \bar\Phi$. Note the relation
\be
\Phi = i\left[X\,u^-_2u^+_2 -\frac{1}{2}\,D^{++}X\,u^-_2u^-_2
-\frac{1}{2}\,D^{--}X\,u^+_2u^+_2\right]. \label{PhiX}
\ee
The free action \p{FreeqCov} can also be expressed through the universal invariant $X$:
\be
S^{free}_{cov} =  -\frac{3}{2} \int \mu_H\,  X^2\,.
\ee
This relation can be proved, starting from the $\tau$ form of the free action \p{ActFreetau} and
integrating by parts with respect to the harmonic derivatives. Another way to see this
is to compare both sides in the original $\lambda$ frame representation
(i.e. in terms of $q^{\pm a}$) by choosing the Wess-Zumino gauge for the superfields $V^{\pm\pm}_1$
and $V^{\pm\pm}_2$
$$
V^{\pm\pm}_I = \theta^\pm\bar\theta^\pm B_I\,.
$$

Finally, we address the issue of the Fayet-Iliopoulos (FI) terms. In the present case one can
define two independent gauge invariant FI terms
\be
S^{FI}_1 = i\xi_1 \int du d \zeta^{(-2)}\, V^{++}_1\,, \quad
S^{FI}_2 = i\xi_2 \int du d \zeta^{(-2)}\, V^{++}_2\,. \label{FILin}
\ee
Let us consider the first one. Rewriting it in the full harmonic superspace
\be
S^{FI}_1 = i\xi_1 \int \mu_H \,\theta^-\bar\theta^- D^{++}v_1 =
-i\xi_1 \int \mu_H \,(\theta^+\bar\theta^- + \theta^-\bar\theta^+)v_1\,,
\ee
expressing $v_1$ from the relation \p{Q+def}, integrating by parts
with using the analyticity property of $q^{+ a}\,$ and $V^{++}_1, V^{++}_2$  and
performing in the end the integration over harmonics, this term
in the gauge \p{GauGe} and frame \p{frame} can be transformed
to the expression
\be
S^{FI}_1 = -\frac{i}{2}\xi_1\,\int dt d^4\theta \,\left(\theta_1\bar\theta^2 \Phi
-\theta_2\bar\theta^1 \bar\Phi\right). \label{FI1}
\ee
It is a particular case of twisted chiral superpotential term. Analogously,
\be
S^{FI}_2 = \frac{1}{2}\xi_2\,\int dt d^4\theta \,\left(\theta_1\bar\theta^2 \Phi +
\theta_2\bar\theta^1 \bar\Phi\right). \label{FI2}
\ee
It is unclear whether a general chiral superpotential can be generated from
some gauge invariant $q^{+ a}$ action.

\setcounter{equation}{0}
\section{Nonlinear chiral multiplet}
Let us now consider the gauging of the two-parameter abelian symmetry (\ref{Abel}b)
\be
\delta q^{+ a} = \Lambda_1 q^{+ a} + \Lambda_2\, c^{a}_{\;\;b}q^{+ b}\,, \quad
\Lambda_{I} = \Lambda_{I}(\zeta, u)\,, \; I=1,2\,. \label{NLqtr}
\ee
The harmonic constraint (\ref{qcond}b) is now covariantized as
\be
D^{++}q^{+ a} - V^{++}_1 q^{+a} - V^{++}_2 c^{a}_{\;\; b}q^{+b} = 0\,. \label{NLqconstr}
\ee
The analytic potentials $V^{++}_I,\; I=1,2\,$, possess the same gauge transformation
laws \p{TranV} and are
expressed through the bridges $v_I$ with the mixed transformation rules \p{TranBrid}
by the same relations
\p{Brid1}. However, since now $q^{+ a}$ transforms homogeneously under the gauge transformations,
the relation \p{Q+def} between the $\lambda$ and $\tau$ world objects has to be modified:
\bea
q^{+a} = e^{v_1}\left(\cos v_2\, Q^{+ a} + \sin v_2 c^{a}_{\; b}\,Q^{+b}\right), \quad
Q^{+a} = e^{-v_1}\left(\cos v_2\, q^{+ a} - \sin v_2 c^{a}_{\; b}\,q^{+b}\right), \label{Qq}
\eea
or, in another form,
\be
q^{+a} + ic^{a}_{\;\;b}\, q^{+b} = e^{v_1 -iv_2}\left(Q^{+a}
+ ic^{a}_{\;\;b}\, Q^{+b}\right), \quad
q^{+a} - ic^{a}_{\;\;b}\, q^{+b} = e^{v_1 +iv_2}\left(Q^{+a}
- ic^{a}_{\;\;b}\, Q^{+b}\right). \label{Qq1}
\ee

A direct calculation shows that the constraint \p{NLqconstr} entails, for $Q^{+ a}\,$,
\be
D^{++} Q^{+ a} = 0\,,\label{QHconstr}
\ee
whence, in the central basis,
\be
Q^{+ a} = Q^{i a}(t,\theta, \bar\theta) u^{+}_i \label{LinQ}
\ee
(cf. \p{DefQ}). Also, using the transformation laws \p{TranBrid} and \p{NLqtr}, it is easy to find
\be
\delta Q^{\pm a} = \tau_1 Q^{\pm a} + \tau_2 c^{a}_{\;\;b}Q^{\pm b}\,. \label{taugauge}
\ee

In what follows it will be convenient to choose the $SU(2)$ frame \p{frame} in which
\be
q^{+ 1} = e^{(v_1 + iv_2)}\,Q^{+ 1}\,, \quad q^{+ 2} = e^{(v_1 - iv_2)}\,Q^{+ 2}\,, \label{q-Q}
\ee
\bea
\delta Q^{11} = (\tau_1 + i\tau_2)\,Q^{11}, \delta Q^{22} = (\tau_1 - i\tau_2)\,Q^{22},
\delta Q^{21} = (\tau_1 + i\tau_2)\,Q^{21},  \delta Q^{12} = (\tau_1 - i\tau_2)\,Q^{12}.\label{tau1}
\eea
Like in the previous Section, the analyticity of $q^{+a}$ implies the ``covariant analyticity'' for
$Q^{+ a}$:
\be
D^{+}q^{+ 1} = \bar D^{+}q^{+ 1} = 0 \, \Leftrightarrow \,
\left[ D^{+} + D^{+}(v_1 + iv_2)\right] Q^{+ 1} =
\left[ \bar D^{+} + \bar D^{+}(v_1 + iv_2)\right] Q^{+ 1} = 0\,, \label{CovAn1}
\ee
\be
D^{+}q^{+ 2} = \bar D^{+}q^{+ 2} = 0 \, \Leftrightarrow \,
\left[ D^{+} + D^{+}(v_1 - iv_2)\right] Q^{+ 2} = \left[ \bar D^{+}
+ \bar D^{+}(v_1 - iv_2)\right] Q^{+ 2} = 0\,. \label{CovAn}
\ee
{}From the analyticity of $V^{++}_{1,2}$  it follows that the gauge connections
in \p{CovAn1}, \p{CovAn} are linear in harmonics
\bea
&& D^{+}(v_1 + iv_2) = A^{a}_{(+)}(t, \theta, \bar\theta)u^{+}_a\,, \quad
D^{+}(v_1 - iv_2) = A^{a}_{(-)}(t, \theta, \bar\theta)u^{+}_a\,, \nn
&& \bar D^{+}(v_1 - iv_2) =  -\bar{A}^{a}_{(+)}(t, \theta, \bar\theta)u^{+}_a\,, \quad
\bar D^{+}(v_1 + iv_2) = -\bar{A}^{a}_{(-)}(t, \theta, \bar\theta)u^{+}_a\,. \label{Conn}
\eea
Now it is time to properly fix the $\tau $ gauge freedom \p{tau1}. Assuming that
$Q^{12} = -\overline{(Q^{21})}$
possesses  a non-zero constant background, a convenient gauge is
\be
Q^{12} = -Q^{21} = 1\,. \label{GaugeNL}
\ee
Substituting this gauge into the covariant analyticity conditions for $Q^{+a}$ in \p{CovAn},
taking into account the relations \p{LinQ}, \p{Conn}, and equating to zero the coefficients
of three independent products of
harmonics ($(u^{+}_1)^2, (u^+_2)^2$ and $u^+_1u^+_2$), we obtain
\bea
&& A^2_{(+)} = A^1_{(-)}= 0\,, \quad A^1_{(+)} = D^2 \Phi\,, \; A^2_{(-)} = -D^1 \bar\Phi \,,
\quad \mbox{(and c.c.)}\,, \nn
&& D^1 \Phi + \Phi D^2 \Phi =  \bar D^1 \Phi + \Phi  \bar D^2 \Phi= 0\,, \quad
D^2 \bar\Phi - \bar\Phi D^1 \bar\Phi = \bar D^2 \bar\Phi - \bar\Phi \bar D^1 \bar\Phi = 0\,,
\label{TwNonl}
\eea
where $Q^{11} \equiv \Phi\,$, $Q^{22} = \bar\Phi \,$.

Thus  the only independent object that remains in the gauge \p{GaugeNL} is a complex
${\cal N}{=}4$ superfield  $\Phi$ subjected to the constraints \p{TwNonl}. These constraints
are a nonlinear version of the twisted chirality constraints \p{ChirTw} and are easily recognized
as a twisted version of the {\it nonlinear ${\cal N}{=}4$ chirality constraints} \cite{IKL}.
It can be given the form of the ordinary nonlinear chirality constraints by relabelling
the covariant derivatives just in the same way as in the case of the linear constraints \p{ChirTw}.

It is worth mentioning a specific feature of the nonlinear chiral multiplet case as
compared with the linear multiplet case. The original $\lambda$ world constraints (\ref{qcond}a),
\p{NLqconstr} preserve the whole automorphism $SU(2)_{R}$ group acting on the doublet indices
of harmonics and Grassmann coordinates and break the Pauli-G\"ursey
$SU(2)_{PG}$ symmetry realized on the doublet index $a$ of $q^{+a}$ down to a $U(1)$ subgroup
(due to the presence of the constant triplet $c^{ab}$ in \p{NLqconstr}). The same symmetry structure
is exhibited by the $\tau$ world constraints \p{QHconstr}, \p{CovAn} which are equivalent
to \p{NLqconstr} and the analyticity condition (\ref{qcond}a). Before fixing
the $\tau$ frame gauge as in \p{GaugeNL}, the superfields $Q^{ia}$ are transformed
by $SU(2)_R$ linearly,
in the same way as the spinor derivatives $D^{i}, \bar D^{i}$, i.e. as
\be
\delta_R Q^{ia} \simeq Q^{ia}{}'(t, \theta', \bar\theta') -  Q^{ia}(t, \theta, \bar\theta) =
\lambda^i_{\;\;k} Q^{ka}\,, \quad \overline{(\lambda^{ik})} = \lambda_{ik}\,, \;\;
\lambda^i_{\;\;i} =0\,.
\ee
Here $\lambda^{ik}$ are constant $SU(2)_R$ parameters. After imposing the gauge \p{GaugeNL},
this transformation law becomes nonlinear, as it must be accompanied by
the compensating $\tau$ gauge transformation needed for preserving \p{GaugeNL}
\be
(\tau_1 - i\tau_2)_{comp} = -\lambda^{12} + \lambda^{11}\, \bar\Phi\,, \quad
(\tau_1 + i\tau_2)_{comp} = \lambda^{12} + \lambda^{22}\, \Phi\,,
\ee
whence
\be
\delta_R \Phi = \lambda^{11} + 2\lambda^{12} \Phi + \lambda^{22}(\Phi)^2 \quad \mbox{and c.c.}\,,
\label{S2int}
\ee
i.e. $\Phi$ and $\bar\Phi$ are transformed as projective $CP^1$ coordinates of
the 2-sphere $S^2 \sim SU(2)_R/U(1)_R$. The constraints \p{TwNonl}, where $D^i$ and $\bar D^i$
are still transformed linearly with respect to their
doublet indices and $\Phi\,$, $\bar\Phi$ are transformed according to the nonlinear
transformation rule \p{S2int}, are directly checked to be $SU(2)_R$ covariant.
This interpretation of the nonlinear chiral ${\cal N}{=}4, d{=}1$ superfields as
parameters of $S^2$ was the starting point of the derivation of the constraints \p{TwNonl}
in \cite{IKL} (in an ${\cal N}{=}4$ superspace parametrization twisted as compared to ours).
Note that both $\Phi$ and $\bar\Phi$ are inert under the $U(1)$ remnant of the broken
$SU(2)_{PG}$ symmetry.

Let us now discuss how the actions of the nonlinear chiral multiplet are reproduced
from the gauged $q^{+}$ actions.

The standard free action of $q^{+ a}$ is obviously not invariant even under the rigid
version of \p{NLqtr} due to the presence of the rescaling isometry in \p{NLqtr}. This situation
is quite similar to what we faced in \cite{1} when deriving the action of the nonlinear
${\bf (3, 4, 1)}$ multiplet from the $q^+$ action with a gauged rescaling
invariance. It was shown there that the simplest invariant action is a nonlinear action
of the sigma-model type. In the case considered here we should start from the action
which is simultaneously invariant under the rescalings and the $U(1)$ transformations.
The unique object invariant under both gauged isometries
is constructed as
\be
Y = \frac{q^{+ a}c_{ab}q^{-b}}{\left(q^{+ a}q^-_{a}\right)}\, ,\label{InvNL}
\ee
where
\be
q^{- a} \equiv D^{--}q^{+ a}  - V^{--}_1 q^{+a} - V^{--}_2 c^{a}_{\, b}q^{+b}
\ee
and $V^{--}_{1,2}$ were defined in \p{V--}. It is the true nonlinear analog of the invariant
$X$ defined in \p{DefX}. It is easy to check that  $Y$ admits an equivalent representation in terms
of the $\tau$ world objects, such that it is  manifestly invariant under the $\tau$ gauge
transformations \p{taugauge}
\be
Y = \frac{Q^{+ a}c_{ab}Q^{-b}}{\left(Q^{+ a}Q^-_{a}\right)}\, ,
\ee
and satisfies the relations
\be
(D^{++})^2Y= (D^{--})^2Y= 0\,, \;\;  D^{++}D^{--}Y = 2Y\,, \;\; (D^{++}Y\,D^{--}Y)
 - Y^2 = 1\,.\label{relY}
\ee

In the gauge \p{GaugeNL} and frame \p{frame}
\be
Y = \frac{2i}{\left(1 + \Phi\bar\Phi\right)}\left[\bar\Phi \,u^+_2u^-_2 -\Phi\, u^+_1u^-_1 +
\frac{1}{2}\left( 1 - \Phi\bar\Phi \right)\left( u^+_1u^-_2 +  u^+_2u^-_1\right)\right].
\ee
A nonlinear analog of the relation \p{PhiX} is
\be
\frac{\Phi}{1 + \Phi\bar\Phi} = i\left(Y\,u^-_2u^+_2  -\frac{1}{2}D^{++} Y\,u^-_2u^-_2
-\frac{1}{2}D^{--} Y\,u^+_2u^+_2\right).\label{PhiY}
\ee
{}From this relation and its conjugate it is easy to express $\Phi$ and $\bar\Phi\,$ through
$Y$ and its harmonic derivatives.

The general sigma-model type action
of the nonlinear chiral multiplet corresponds to the following gauged subclass
of the general $q^{+ a}$ actions
\be
S_{chn} = \int \mu_H\,{\cal L}(u, Y,D^{++} Y, D^{--} Y)\,. \label{GenNL}
\ee
After expressing $Y$ in terms of the $\tau$ world objects, choosing
the gauge \p{GaugeNL}, $SU(2)$ frame \p{frame} and preforming the integration over harmonics,
\p{GenNL} becomes the general sigma model action of the nonlinear chiral multiplet $\Phi, \bar\Phi$
as it was given in \cite{BBKNO}.

Let us point out that this correspondence, as in other similar cases \cite{1,2}, allows one
to equivalently deal with the action in the original $\lambda$ frame $q^{+}$ representation
by choosing the appropriate Wess-Zumino gauge for the ``topological'' gauge superfields $V^{\pm\pm}_{1,2}$
and using the residual two-parameter gauge freedom to trade two out of four original physical bosonic
fields of $q^{+ a}\,$ for two $d{=}1$ ``gauge fields''. The latter become just two auxiliary fields of the
nonlinear chiral multiplet. To work in the $q^{+}$ representation is in some aspects easier than to use
the $\tau$ frame where the same action looks as the general action of the superfields
$\Phi$ and $\bar\Phi\,$.

It is interesting that in the case under consideration there also exists a unique $SU(2)_R$ invariant
action of the WZW type which is gauge invariant up to a total derivative in the integral and
in this sense is an analog of the free action of the linear chiral multiplet in the form \p{FreeqCov},
\p{rontudju}. The quantity
\be
F = q^{+a}q_a^-  = e^{2v_1} Q^{+a}Q^-_a
\ee
is manifestly $SU(2)_R$ invariant and invariant under the gauge $\Lambda_2$ transformations, while its
$\Lambda_1$ transformation reads
\be
\delta_{\Lambda_1} F = 2\Lambda_1\,F\,, \quad \delta_{\Lambda_1} \log F = 2 \Lambda_1\,.
\ee
Then the action
\be
S_{su(2)} = \int \mu_H \, \log F = \int \mu_H \, \left[2v_1 + \log (Q^{+a}Q^-_a)  \right] \label{su2}
\ee
is gauge invariant since the full ${\cal N}{=}4$ superspace integral of the analytic parameter $\Lambda_1$
vanishes. Using \p{Qq1}, it is easy to check that the bridge $v_1$, modulo a constant
and purely analytic term
which vanish after integration, in the gauge \p{GaugeNL} is reduced to
\be
v_1 \,\Rightarrow \, \log \left(1 - u^+_1 u^-_2 + 2\Phi u^+_1 u^-_1 \right) +
\log \left(1 + u^+_2 u^-_1 + 2\bar\Phi u^+_2 u^-_2 \right).
\ee
It follows from the constraints \p{TwNonl} that the integral of any holomorphic or antiholomorphic
function of $\Phi, \bar\Phi$ over the full ${\cal N}{=}4$ superspace vanishes. So the bridge $v_1$
drops out from  \p{su2} in the gauge \p{GaugeNL} and, taking into account that in this gauge
\be
Q^{+a}Q^-_a = 1 + \Phi\bar\Phi\,,
\ee
we obtain the simple final expression for the action \p{su2}
\be
S_{su(2)} = \int dtd^4\theta \, \log (1 + \Phi\bar\Phi). \label{su2a}
\ee
It describes the ${\cal N}{=}4$ superextension of the $d{=}1$ sigma model on $S^2 \sim SU(2)_R/U(1)_R$
and was derived in this form for the first time in \cite{IKL}. The Lagrangian in \p{su2a}
is just the corresponding
K\"ahler potential \footnote{Formally, \p{su2a} looks also invariant under $SU(2)_{PG}$, however the
constraints \p{TwNonl} do not respect this second $SU(2)$, so the latter is not a symmetry
of \p{su2a}. }. Actually, using eq. \p{PhiY}, we could write an equivalent representation for $S_{su(2)}$
through the universal tensorial invariant $Y$. However, the $SU(2)_R$ symmetry in such a representation
is not manifest due to the presence of explicit harmonics over which one should integrate.

Finally, we discuss the structure of two FI terms in the present case. They are originally given by the
same $V^{++}_{1, 2}$ actions \p{FILin} as in the linear case.
Further, passing to the integral over the full ${\cal N}{=}4$ superspace, using the relation
\be
v_1 -iv_2 = \log \left[(q^{+ a} + i c^{a}_{\;\; b} q^{+ b})u^-_a\right] -
\log \left[(Q^{+ a} + i c^{a}_{\;\; b} Q^{+ b})u^-_a\right]
\ee
and its conjugate (they follow from \p{Qq1}), as well as the analyticity of $q^{+ a}$, and,
finally, performing the integration over harmonics in the gauge \p{GaugeNL}, the FI terms
can be expressed through $\Phi$, $\bar\Phi$ as follows
\be
\tilde{S}^{FI}_1 = -\frac{i}{2}\tilde{\xi}_1\,\int dt d^4\theta \,\left(\theta_1\bar\theta^2 \Phi
-\theta_2\bar\theta^1 \bar\Phi\right), \quad \tilde{S}^{FI}_2
= \frac{1}{2}\tilde{\xi}_2\,\int dt d^4\theta \,\left(\theta_1\bar\theta^2 \Phi +
\theta_2\bar\theta^1 \bar\Phi\right).
 \label{FINL}
\ee
Surprisingly, they are still linear in $\Phi $, $\bar\Phi$, like their customary chiral superfield
analogs \p{FI1}, \p{FI2}. Note that the $SU(2)_R$ invariance of \p{FINL} can be checked with the help of the
basic constraints \p{TwNonl}, taking into account that $SU(2)_R$ transforms the explicit $\theta$s
in  \p{FINL} in the standard way (rotates them in the doublet index), while $\Phi$ and $\bar\Phi$ are
transformed according to the law \p{S2int} and its conjugate.

\setcounter{equation}{0}
\section{Nonabelian gauge group}
Let us now consider gauging of the last two-parameter group admitting a realization on the analytic
superfield $q^{+ a}$, the non-abelian solvable group \p{Nonabel} consisting of a dilatation and a shift
of $q^{+a}\,$. The gauge transformation laws are
\begin{equation}
\delta q^{+a}=\Lambda_{1}q^{+a}+\Lambda_{2}u^{+a}\,, \label{Nonabel1}
\end{equation}
where as before $\Lambda_{1}$ and $\Lambda_{2}$ are charge-zero analytic superfields. Due to the nonabelian
character of this group, its gauging is a little bit more tricky as compared to the previous
two (abelian) cases. The commutation relations of the gauge transformations are given by
\begin{equation}
[\delta,\delta']q^{+a}=\delta''q^{+a}\,,\quad \Lambda''_{1}=0\,,\,\,\,\Lambda''_{2}
=\Lambda_{2}\Lambda'_{1}-\Lambda_{1}\Lambda'_{2}\,. \label{Noncomm}
\end{equation}
In order to covariantize the harmonic constraints, we need to introduce two analytic gauge
superfields $V^{++}$, $W^{++}$ with the transformation laws
\begin{equation}
\delta V^{++}=D^{++}\Lambda_{1}\,,\quad
\delta W^{++}=D^{++}\Lambda_{2}+\Lambda_{1}W^{++}-\Lambda_{2}V^{++}.
\end{equation}
It is easy to check that the Lie bracket of two such transformations has the form \p{Noncomm}.
The covariant harmonic constraint on the superfield $q^{+a}$ now reads
\begin{equation}
D^{++}q^{+a}-V^{++}q^{+a}-W^{++}u^{+a}=0\,.\label{cons01}
\end{equation}
The gauge superfields $V^{++}$ and $W^{++}$ are expressed through the corresponding non-analytic
bridge superfields $v$, $w$ as
\begin{equation}
V^{++}=D^{++}v\,,\quad W^{++}=e^vD^{++}w\,.
\end{equation}
As in other cases, the bridges $v$, $w$ transform under two types of gauge transformations,
the original ones with the analytic parameters $\Lambda_{1}(\zeta,u)$ and $\Lambda_{2}(\zeta,u)$,
and new ones with the parameters $\tau_{1}(t,\theta, \bar\theta)$ and $\tau_{2}(t,\theta, \bar\theta)$
which are independent of harmonic variables:
\begin{equation}
\delta v =\Lambda_{1}(\zeta,u)-\tau_{1}(t,\theta, \bar\theta)\,,\quad
\delta w =e^{-v}\,\Lambda_{2}(\zeta,u)- \tau_{2}(t,\theta, \bar\theta) +\tau_{1}(t,\theta, \bar\theta)\,w\,.
\end{equation}
We also define the new non-analytic ``$\tau$-world'' superfield
\begin{equation}
Q^{+a}=e^{-v}q^{+a}-wu^{+a}\,,\quad \delta Q^{+a}= \tau_{1}Q^{+a} +\tau_{2}u^{+a}\,.
\end{equation}
As a consequence of (\ref{cons01}), the superfield $Q^{+a}$ is homogeneous in the harmonic variables
\begin{equation}
D^{++}Q^{+a}=0\,\,\Rightarrow\,\, Q^{+a}(t,\theta, \bar\theta, u)
=Q^{ba}(t,\theta, \bar\theta)u^+_{b}\,, \;\;
\overline{(Q^{ab})} = Q_{ab}\,.
\end{equation}
The harmonic-independent superfields $Q^{ba}$ transform as
\begin{equation}
\delta Q^{ba}= \tau_{1}Q^{ba} -\tau_{2}\epsilon^{ba}.
\end{equation}
The real parameter $\tau_{2}$ may be used to gauge away the antisymmetric part of the tensor $Q^{ba}$:
\be
Q^{ab} = Q^{(ab)}\,. \label{Gauge0}
\ee
Then the target space scale invariance with the real parameter $\tau_{1}$ may be used to fix the
value of a component of this tensor. We choose the gauge
\begin{equation}
Q^{21}=Q^{12}=i\,\,\Rightarrow\,\, Q^{+1}=Q^{11}u^+_{1}+iu^+_{2}\,,\quad
Q^{+2}=Q^{22}u^+_{2}+iu^+_{1}\,.\label{Gauge3}
\end{equation}
We now should take into account the analyticity constraints on the original superfields $q^{+a}$.
When expressed in terms of the new superfields, these constraints become
\begin{equation}\begin{array}{l}
D^+Q^{+a}+D^+v\,Q^{+a} +(D^+w+wD^+v)\,u^{+a}=0\,,\cr \bar D^+Q^{+a}+\bar D^+v\,Q^{+a}
+(\bar D^+w+w\bar D^+v)\,u^{+a}=0\,.\end{array}\label{ralbo1}
\end{equation}
Due to the analyticity of the gauge superfields $V^{++}$, $W^{++}$, the fermionic connections
$D^+v$, $D^+w+wD^+v$ have a simple dependence on harmonic variables
\begin{eqnarray}
D^{++}D^+v &=& 0 \quad \Rightarrow \quad D^+v =A^{a}(t,\theta,\bar\theta)u^+_{a}, \label{Aconn} \\
D^{++}(D^+w+wD^+v) &=& 0 \quad \Rightarrow \quad D^+w+w\,D^+v=B^{a}(t,\theta,\bar\theta )u^+_{a}\,.
\label{Bconn}
\end{eqnarray}
Analogously, for the conjugate connections we have
\begin{equation}
\bar D^+v= -\bar A^{a}u^+_{a}\,,\quad \bar D^+w+w\bar D^+v= -\bar B^{a}u^+_{a}\,.
\end{equation}
Then, the constraints (\ref{ralbo1}) imply, in ordinary superspace,
\bea
&& D^2Q^{11}+A^2Q^{11}+iA^1-B^1=0\,, \quad B^2=iA^2 \,, \nn
&& D^1Q^{22}+A^1Q^{22}+iA^2+B^2=0\,, \quad B^1=-iA^1\,, \label{AB} \\
&& D^1Q^{11}+A^1Q^{11}=0\,, \quad D^2Q^{22}+A^2Q^{22}=0\,.\label{Constr1}
\eea
{}From eq. \p{AB} one expresses the gauge connections
\begin{eqnarray}
&& A^1=iB^1=\frac{1}{4+Q^{11}Q^{22}}\left(2iD^2Q^{11}-Q^{11}D^1Q^{22}\right), \nn
&& A^2=-iB^2=\frac{1}{4+Q^{11}Q^{22}}\left(2iD^1Q^{22}-Q^{22}D^2Q^{11}\right). \label{Aexpr}
\end{eqnarray}
Substituting these expressions into eqs. \p{Constr1} and then using complex conjugation
yield the full set of the nonlinear constraints on the superfields $Q^{11} \equiv \Phi\,,
Q^{22} = \bar\Phi$
\begin{eqnarray}
&& D^1\Phi +\frac{\Phi}{4+ \Phi\bar\Phi}\left(2iD^2\Phi - \Phi D^1\bar\Phi\right)=0\,,
\; \bar D^1\Phi+\frac{\Phi}{4+ \Phi \bar\Phi}\left(2i\bar D^2\Phi - \Phi\bar D^1\bar\Phi\right)=0\,,
\label{ralbo2}\\
&& D^2\bar\Phi+\frac{\bar\Phi}{4+ \Phi\bar\Phi}\left(2iD^1\bar\Phi- \bar\Phi D^2\Phi\right)=0\,, \;
\bar D^2\bar\Phi+\frac{\bar\Phi}{4+ \Phi\bar\Phi}\left(2i\bar D^1\bar\Phi
- \bar\Phi \bar D^2\Phi\right)=0\,.
\label{ralbo3}
\end{eqnarray}
Equations (\ref{ralbo2}) and (\ref{ralbo3}) may be interpreted as (twisted) non-linear chirality
constraints on the superfields $\Phi$ and its complex conjugate $\bar\Phi\,$.

Let us now define the subclass of the general $q^{+}$ actions which respects the invariance
under the rigid transformations \p{Nonabel} and, after gauging, under their local counterparts
\p{Nonabel1}. In the $\tau$ frame it should yield the general sigma-model type action of
the nonlinear chiral multiplet in question.

The invariance under the shift transformations in \p{Nonabel} just means that the corresponding
superfield Lagrangian cannot depend  on the trace part in $q^{ia}u^{+}_i$, i.e.
$q^{ia} \rightarrow q^{(ib)}\,$.
Next, the invariance under the target space scale transformations constrains the action to depend
only on two independent ratios of three components  of
$q^{(ab)}$, i.e. $q^{(ab)} \rightarrow q^{11}/q^{12}, \, q^{22}/q^{12}\,$. In other words,
the appropriate general superfield $q^{+ a}$ Lagrangian should be an arbitrary function of
the superfields $q^{11}/q^{12}, \, q^{22}/q^{12}\,$ which can be interpreted as projective
coordinates of some two-sphere $S^2$ . Being reformulated in terms of the superfields $q^{\pm a}$,
this requirement amounts to the following  particular choice of the $q^+$ lagrangian
\be
{\cal L} = {\cal L}\left(u^{\pm}, \frac{\hat{q}{}^{\pm a}}{\vert \hat{q}\vert}\right), \label{subcl1}
\ee
where
\bea
&& \hat{q}{}^{\pm a} =  q^{\pm a} -\frac{1}{2}u^{\pm a}\left(q^{+ b}u^-_b - q^{- b}u^+_b \right), \; \nn
&& \vert \hat{q}\vert^2 = \hat{q}{}^{+ a}\hat{q}{}^{-}_{a} = (q^{+ a}u^+_a)(q^{- b}u^-_b) -
\frac{1}{4}(q^{+ a}u^-_a + q^{- a}u^+_a)^2 =
\frac{1}{2}\, q^{(ab)}q_{(ab)}\,. \label{Bas2}
\eea
As in the previous case, the standard free action of $q^{+ a}$ is not invariant under the target
space rescalings. The Lagrangians from the subclass \p{subcl1} are of the sigma-model type,
with non-constant bosonic target metrics.

The gauging of these Lagrangians goes in the standard way, by subjecting $q^{+ a}$ to the covariantized
constraint \p{cons01} and defining $q^{- a}$ in a gauge-covariant way as
\begin{equation}
q^{-a}=D^{--}q^{+a}-V^{--}q^{+a}-W^{--}u^{+a}\,,\quad \delta q^{-a}=\Lambda_{1}q^{-a}+\Lambda_{2}u^{-a}\,.
\end{equation}
The final gauge-covariantized action has the same form as the rigidly invariant
one \p{subcl1} but with the superfields $q^{\pm a}$ defined in a gauge-covariant way. Just due
to this covariance, the basic objects \p{Bas2} admit the equivalent $\tau$ frame representation
\bea
&& \hat{q}{}^{\pm a} =  e^{v}\left[ Q^{\pm a} -\frac{1}{2}u^{\pm a}\left(Q^{+ b}u^-_b
- Q^{-b}u^+_b \right)\right]
\equiv e^v\,\hat{Q}{}^{\pm a} , \nn
&& \vert \hat{q}\vert^2 = e^{2v}\left[(Q^{+ a}u^+_a)(Q^{- b}u^-_b) - \frac{1}{4}\left(Q^{+ a}u^-_a
 + Q^{- a}u^+_a\right)^2\right] \nn
&& \quad \;\;\,=
\frac{1}{2}\,e^{2v} Q^{(ab)}Q_{(ab)} \equiv e^{2v}\,\vert \hat{Q}\vert^2 \,. \label{Bas3}
\eea
The covariantized superfield argument in \p{subcl1} does not depend on the bridge $v\,$, whence
\be
{\cal L} = {\cal L}\left(u^{\pm}, \frac{\hat{q}{}^{\pm a}}{\vert \hat{q}\vert}\right) =
{\cal L}\left(u^{\pm}, \frac{\hat{Q}{}^{\pm a}}{\vert \hat{Q}\vert}\right). \label{subcl12}
\ee
In the gauges \p{Gauge0} and \p{Gauge3}:
\be
\hat{Q}{}^{\pm 1} = {Q}^{\pm 1}  = \Phi u^\pm_{1}+iu^\pm_{2}\,, \;
\hat{Q}{}^{\pm 2} = {Q}^{\pm 2}  = \bar\Phi u^\pm_{2}+iu^\pm_{1}\,, \quad \vert \hat{Q}\vert
= \sqrt{1 + \Phi\bar\Phi}\,,  \label{Qphi5}
\ee
and the action corresponding to the Lagrangian \p{subcl12}, after performing the integration over
harmonics, becomes the general off-shell action of the supermultiplet $\Phi$, $\bar\Phi$.
Note that the relations \p{Qphi5}, like analogous relations of the previous cases,  are invertible:
\be
\Phi = \hat{Q}{}^{+ 1}u^{-}_2 - \hat{Q}{}^{- 1}u^{+}_2\,, \quad \bar\Phi =
\hat{Q}{}^{- 2}u^{+}_1 - \hat{Q}{}^{+ 2}u^{-}_1\,.
\ee
This ensures the possibility to express $\Phi, \bar\Phi $ through the basic gauge invariant object,
$\hat{q}{}^{\pm a}/\vert \hat{q}\vert = \hat{Q}{}^{\pm a}/\vert \hat{Q}\vert\,$,
and in fact proves the equivalence of the general ${\cal N}{=}4$ action of superfields
$\Phi, \bar\Phi$ and
the particular class of the gauged $q^{+}$ actions defined above.

It is worth noting that the building blocks of the $\lambda$ world gauge invariants
can be successively
reproduced from the simplest invariant of the shift $\lambda$ gauge transformation
(with the parameter $\Lambda_2$)
\begin{equation} q^{++}=q^{+a}u^+_{a} = \hat{q}{}^{+ a}u^+_a\,,
\quad \delta q^{++}=\Lambda_{1}q^{++}\,.
\label{q++}
\end{equation}
Acting on \p{q++} by the covariant derivative $D^{--} - V^{--}$, we can
produce new non-analytic superfields
which are invariant under the $\Lambda_2$ transformations and covariant with respect
to the $\Lambda_1$ transformations:
\begin{equation} q^{+-}=\frac{1}{2}(D^{--}-V^{--})q^{++},\,\,
q^{--}=(D^{--}-V^{--})q^{+-},\,\,\delta
q^{\pm -}=\Lambda_{1}q^{\pm -}\,.
\end{equation}
They are related to the superfields $q^{\pm a}$ and $\hat{q}{}^{\pm a}$ by
\begin{equation}
q^{+-}=\frac{1}{2}(q^{+a}u^-_{a}+q^{-a}u^+_{a}) = \hat{q}{}^{+ a}u^-_a
= \hat{q}{}^{- a}u^+_a \,,\,\,\,
q^{--}=q^{-a}u^-_{a} = \hat{q}{}^{- a}u^-_a
\end{equation}
and can be used to form two independent gauge invariant ratios
\begin{equation}
X^{++}=\frac{q^{++}}{\sqrt{q^{++}q^{--}-( q^{+-})^2}}\,,
\quad X^{--}=\frac{q^{--}}{\sqrt{q^{++}q^{--}-( q^{+-})^2}}\,,
\end{equation}
which are just independent harmonic projections of the superfield argument in \p{subcl12}.

Let us now dwell on the peculiarities of the realization of $SU(2)_R$
and $SU(2)_{PG}$ symmetries on
the superfields $\Phi$ and $\bar\Phi$ and the surprising relation to the nonlinear chiral multiplet
discussed in the previous Section.

The basic gauge covariant constraint \p{cons01} clearly breaks the original
$SU(2)_R\times SU(2)_{PG}$ symmetry realized on $q^{+ a}$, Grassmann and harmonic coordinates down
to the diagonal $R$-symmetry group $SU(2){}'_R$  which uniformly rotates all doublet indices.
The gauge \p{Gauge0} is  $SU(2){}'_R$ covariant, so the superfield $Q^{(ab)}$ is
transformed as
\be
\delta_{R'} Q^{(ab)} \simeq Q^{(ab)}{}'(t, \theta',u') - Q^{(ab)}(t,\theta, u) =
\lambda^{a}_{\;\;d}Q^{(db)} + \lambda^{b}_{\;\;d}Q^{(ad)}\,,
\quad  \lambda^{b}_{\;\;b} =0\,.\label{SU2prime}
\ee
The gauge \p{Gauge3} is not preserved under \p{SU2prime}, and in order to restore this gauge
one should accompany the $SU(2){}'_R$ transformations by a compensating $\tau_1$ transformation with
\be
(\tau_1)_{comp} = i\left(\lambda^{22}\Phi - \lambda^{11}\bar\Phi\right).
\ee
As a result, in this gauge the superfields $\Phi$ and $\bar\Phi$ are nonlinearly transformed
under $SU(2){}'_R$
\be
\delta_{R'} \Phi = 2\lambda^{12}\Phi -i\lambda^{11}\left(2 + \Phi\bar\Phi \right)
+ i\lambda^{22} (\Phi)^2\,, \quad
\delta_{R'}\bar\Phi = \overline{(\delta_{R'} \Phi )}\,,
\ee
and so can be treated as coordinates of the coset $S^2 \sim SU(2){}'_R/U(1){}'_R$
in a particular parametrization. Obviously, there should exist an equivalence transformation
to the stereographic projection parametrization in which the $S^2$ coordinates are transformed
according to the holomorphic law \p{S2int}. The precise form of this field redefinition
is as follows
\bea
&& \chi = i\frac{\Phi}{1 + \sqrt{1 + \Phi\bar\Phi}}\,, \quad \Phi = -2i\frac{\chi}{1 - \chi\bar\chi}\,,
\label{redef1}\\
&& \delta_{R'}\chi = \lambda^{11} + 2\lambda^{12} \chi + \lambda^{22}(\chi)^2 \quad \mbox{and c.c.}\,.
\label{chiTran}
\eea
The transformation law \p{chiTran} coincides with \p{S2int}, which suggests that in this new
holomorphic parametrization the constraints \p{ralbo2}, \p{ralbo3} take the form \p{TwNonl}.
Indeed, a simple calculation shows that after the field redefinition \p{redef1} the constraints
\p{ralbo2}, \p{ralbo3} are equivalently rewritten as
\be
D^1 \chi + \chi D^2 \chi = 0\,, \quad \bar D^1 \chi + \chi  \bar D^2 \chi= 0
\quad (\mbox{and c.c.})\,.
\label{chiC}
\ee

Thus we see that the nonlinear chiral multiplet considered in this Section is in fact a disguised form
of the  nonlinear (twisted) chiral multiplet of ref.\cite{IKL} rederived within
the gauging procedure in the previous Section. This is rather surprising, because
in the two cases we gauged two essentially different
two-parameter groups, respectively, abelian and non-abelian ones (\ref{Abel}b) and \p{Nonabel}.
The identity of these two multiplets amounts to the identity of their general actions, despite
the fact that the classes of the appropriate $q^{+}$ actions one starts with in these two cases
are essentially different. Here we again encounter the phenomenon of non-uniqueness of the inverse
oxidation procedure as compared with the target space dimensional reduction \cite{1,2}:
the same off-shell multiplet can be recovered by gauging some non-equivalent isometries
of the ``root'' multiplet. For instance, the ${\cal N}{=}4, d{=}1$ multiplet ${\bf (1, 4, 3)}$
and its most general action can be obtained from the $q^{+}$ multiplet and the appropriate set
of the $q^{+}$ actions by gauging either the non-abelian $SU(2)_{PG}$ group or the abelian group
of three independent shift isometries of $q^{+a}\,$ \cite{2}. Basically, the difference between these
two gauging procedures lies only in the fact that they start from different subclasses
of the general set of $q^{+}$ actions. However, the final action of the
reduced multiplet does not ``remember'' from which parent $q^{+}$ action it originated.

Taking for granted that all  off-shell ${\cal N}{=}4, d{=}1$ superfields
can be recovered from the $q^{+ a}$ superfield
by gauging different symmetries realized on the latter and taking into account
that only three independent
two-parameter groups (defined in \p{Abel} and \p{Nonabel}) can be implemented on $q^{+ a}$,
we conclude that only two essentially different off-shell
${\cal N}{=}4, d{=}1$ multiplets with the content
${\bf (2, 4, 2)}$ exist: the standard linear chiral multiplet
and the nonlinear chiral multiplet introduced in \cite{IKL}.
Any other version of the chiral multiplet should be reducible
to one of these two via some field redefinition.

There is one more way to see that the constraints \p{ralbo2}, \p{ralbo3} are equivalent to \p{TwNonl}.
After some algebra, using \p{ralbo2}, \p{ralbo3} at the intermediate steps, the expressions
for the spinor connections \p{Aexpr} can be cast in the following form
\bea
&& A^1 = D^2\left(\frac{i \Phi}{1 + \sqrt{1 + \Phi\bar\Phi}}  \right)
- D^1 \log \left(1 + \sqrt{1 + \Phi\bar\Phi} \right), \nn
&& A^2 = D^1\left(\frac{i \bar\Phi}{1 + \sqrt{1 + \Phi\bar\Phi}}  \right)
- D^2 \log \left(1 + \sqrt{1 + \Phi\bar\Phi} \right). \label{NewA}
\eea
Substituting these expressions and their complex conjugates into \p{Constr1} and complex
conjugates of \p{Constr1}, we recover \p{chiC}, with $\chi$ being related to $\Phi$
just by eqs. \p{redef1}.

Finally, as an instructive example, we present the $SU(2){}'_R$ invariant action
in terms of the original superfield
variables, as well as the relevant FI term.

The $SU(2){}'_R$ invariant action is given by an expression similar to \p{su2}
\be
S_{su(2)'} = \int \mu_H\, \log \vert \hat{q}\vert =\int \mu_H\, \left(v +
\log \vert \hat{Q}\vert\right). \label{SU22}
\ee
It is manifestly invariant under the gauge shift $\Lambda_2$ transformation (since $\hat{q}{}^{+ a}$ is
invariant), as well as under the scale and shift $\tau$ gauge transformations. It is also invariant
under the scale gauge $\Lambda_1$  transformations since under the latter the Lagrangian in \p{SU22}
is shifted by an analytic gauge parameter the integral of which over the full
${\cal N}{=}4$ superspace vanishes:
\be
\delta_1 \log \vert \hat{q}\vert = \Lambda_1\,, \quad \int \mu_H\,\Lambda_1 = 0\,.
\ee
To find the precise form of the action in terms of the nonlinear chiral superfields $\Phi, \bar\Phi$,
we should make use of eq. \p{Qphi5} and also compute the bridge part of the ${\cal N}{=}4$
superspace integral in \p{SU22}:
\be
\int \mu_H\, v\,.
\ee
This integral can be evaluated by taking one spinor derivative, say $D^{+}$, off the measure $d^4\theta$,
throwing it on $v$, expressing $D^{+}v$ as in \p{Aconn}, doing the harmonic integral $du$,
substituting the gauge-fixed expressions \p{NewA} for the spinor connections $A^1, A^2$
and, finally, restoring the full Grassmann measure
by taking the spinor derivatives $D^1, D^2$ off these expressions. It turns out
that only the second terms
in the expressions \p{NewA} contribute, and we obtain
\be
\int \mu_H\, v\,= - \int dt d^4\theta\, \log \left(1 + \sqrt{1 + \Phi\bar\Phi} \right).
\ee
Using this in \p{SU22}, we obtain
\be
S_{su(2)'} =
\int dt d^4\theta\, \left[\log\,\sqrt{1 + \Phi\bar\Phi} -
\log \left(1 + \sqrt{1 + \Phi\bar\Phi}\right) \right]. \label{SU223}
\ee
Now it is straightforward to check that, after passing to the superfields
$\chi, \bar\chi$ via \p{redef1},
the Lagrangian in \p{SU223} is reduced (modulo a constant shift) just to
$$
\log  \left(1 + \chi\bar\chi\right).
$$
Thus we obtain the expected result that the action \p{SU22}, \p{SU223} is in fact
identical to the previously
considered $SU(2)_R$ invariant action \p{su2a}.

As for the FI terms, in the present case only the gauge superfield
$V^{++}$ possesses an abelian gauge transformation law, so one is able to construct
only one FI term:
\be
S^{FI}_v = i\xi_v \int du d \zeta^{(-2)}\, V^{++} = -i\xi_v \int \mu_H \,(\theta^+\bar\theta^-
+ \theta^-\bar\theta^+)v\,. \label{FI6}
\ee
Inserting the identities $ 1 = D^{+}\theta^{-}\,$, $1 = -\bar{D}{}^+\bar{\theta}{}^-\,$
into the round brackets in the r.h.s. of \p{FI6},  integrating by parts with
respect to spinor derivatives, using the relations \p{Aconn} with \p{NewA}
and their conjugates, doing harmonic integral and, at the end, integrating
by parts once again, one finally finds
\be
S^{FI}_v = -i\xi_v \int dt d^4\theta \,(\theta_1\bar\theta^2 \chi
- \theta_2\bar\theta^1\bar\chi)\,,  \label{FI7}
\ee
which coincides with one of the FI terms in \p{FINL}.

\setcounter{equation}{0}
\section{Conclusions}
In this article and two previous papers \cite{1,2} we showed that all
known off-shell ${\cal N}{=}4, d{=}1$ multiplets
with 4 physical fermions can be  reproduced from the basic (``root'')
multiplet ${\bf(4, 4, 0)}$ by gauging
some symmetries, abelian or non-abelian, realized on this multiplet.
The corresponding general
${\cal N}{=}4$  mechanics actions are recovered as the result of the
proper gauge-fixing in the appropriate gauged subclasses of the general $q^+$ action,
the subclasses which enjoy invariance under the symmetries just mentioned.
Our gauging procedure uses the manifestly supersymmetric universal
language of ${\cal N}{=}4$ superspace and does not require to resort
to component considerations at all. Another merit of our approach is that
it reduces the whole set of non-equivalent superfield actions of
the ${\cal N}{=}4$ mechanics models to some particular cases
of the generic $q^{+}$ action extended by non-propagating ``topological''
gauge superfields. Just the presence of the latter enables one
to preserve the manifest supersymmetry at each step and to reveal
the irreducible off-shell superfield contents of one or another model
by choosing the appropriate superfield gauges and (in the cases
considered in the present paper) by passing to the equivalent $\tau$
frame formulations. The alternative (and in many cases more
technically feasible) way of doing suggested by
the gauging approach is to always stay in the initial $q^{+}$
representation where the harmonic analyticity is manifest
and to choose the WZ gauge for the relevant analytic non-propagating
gauge superfields. Each
``topological'' gauge multiplet in the WZ gauge contributes just
one scalar (``gauge'') field which,
after fully fixing the residual gauge freedom, becomes an auxiliary
field of the new off-shell ${\cal N}{=}4$ multiplet
related to the $q^{+}$ multiplet via linear or nonlinear versions
of the ``automorphic duality'' \cite{ADu}. Thus in the component
formulation our approach automatically yields the explicit
realization of this intrinsically
one-dimensional off-shell duality. The distinctions between various
types of this duality are
related to the differences between the global symmetry groups
subjected to gauging.

The basic peculiarity of the cases considered in this paper as compared
to those treated in \cite{1,2} is that
the superfields describing the ${\cal N}{=}4$ multiplets ${\bf(2, 4, 2)}$
do not ``live'' on the
${\cal N}{=}4$ analytic harmonic subspace (as distinct from the
multiplets  ${\bf(0, 4, 4)}$, ${\bf(1, 4, 3)}$
and ${\bf(3, 4, 2)}$). They are most naturally described after
passing to the equivalent
``$\tau$ frame'' \cite{HSS,HSS1},  with the ordinary ${\cal N}{=}4$
superfield gauge parameters
and the harmonic superfield bridges to the ``$\lambda$ frame''
as the basic gauge objects.
These bridges ensure the equivalence of the manifestly analytic $\lambda$
frame picture one starts with and
the picture in the $\tau$ frame. In the $\tau$ frame, the original
gauge-covariantized analyticity-preserving
harmonic constraints on the superfield $q^{+a}$ amount to the harmonic
independence of the involved superfields.
The harmonic Grassmann analyticity, which is manifest in the $\lambda$ frame, in the
$\tau$ frame amounts to the covariant analyticity conditions. After properly fixing
${\cal N}{=}4$ supersymmetric $\tau$ gauges, these conditions become
the linear or nonlinear ${\cal N}{=}4$
chirality conditions, depending on which two-parameter symmetry group
realized on $q^{+ a}$ is subjected to
gauging. There are only three such groups and they are listed
in \p{Abel} and \p{Nonabel}.
We considered gauging of all these three groups and found that
the gauging of the group (\ref{Abel}a) leads to
the linear chiral ${\cal N}{=}4$ multiplet, while gaugings
of (\ref{Abel}b) and \p{Nonabel} lead to the same
nonlinear chiral multiplet \cite{IKL}, despite the obvious
non-equivalence of these two groups.
This non-uniqueness is a manifestation of the general
non-uniqueness of the oxidation
procedure as inverse to the automorphic duality. Since
only three two-parameter symmetries can be
realized on $q^{+ a}$, from our results it follows,
in particular,  that no other non-equivalent
nonlinear chiral ${\cal N}{=}4$ multiplet can be defined.

Interesting venues for further applications of our gauge approach are
provided by models of ${\cal N}{=}8$
supersymmetric mechanics (see e.g. \cite{ABC,ILS} and refs. therein).
It was argued  in
\cite{ILS}, by considering a wide set of examples,  that
the off-shell ${\cal N}{=}8$ multiplet
${\bf(8, 8, 0)}$ is the true ${\cal N}{=}8$ analog of
the ``root'' ${\cal N}{=}4$ multiplet ${\bf(4, 4, 0)}$
and that the whole set of the component actions of
the ${\cal N}{=}8$ mechanics models with 8 physical
fermions (and finite numbers of auxiliary fields)
follow from the general action of this
basic ${\cal N}{=}8$ multiplet via a linear version
of the automorphic duality. It would be interesting
to apply our techniques to these cases. Recall that
our approach is bound by the requirement that
the symmetries to be gauged commute with supersymmetry.
In the ${\cal N}{=}8$ case
the target space scale and shift transformations still
obey this criterion, so one can hope that the
gauging would nicely work in this case too and could
help to understand the relationships between
the multiplet ${\bf(8, 8, 0)}$  and the rest of
the ${\cal N}{=}8$ multiplets in a manifestly ${\cal N}{=}4$
supersymmetric superfield fashion. We can also hope
to discover in this way new nonlinear ${\cal N}{=}8$
multiplets and the corresponding new ${\cal N}{=}8$ mechanics models,
besides those
already known \cite{N8nl,N8nl2}. The primary question
to be answered is how to define an ${\cal N}{=}8$ analog of
the ${\cal N}{=}4$ topological gauge multiplet which plays a
crucial role in our approach. Another possible way of
extending our study  is to construct ``topological''
${\cal N}{=}4, d{=}1$ supergravity multiplets and
to gauge, with their help, the $R$-symmetry $SU(2)$
groups of ${\cal N}{=}4$ supersymmetry, with new models of
${\cal N}{=}4$ mechanics as an outcome. Finally, let us
note that the nonlinear chiral multiplets
exist also in dimensions $d{>}1$ \cite{IKL},
e.g. in $d{=}3$ \cite{BKS2}. It would be interesting to inquire
whether they can also be derived by gauging some
symmetries realized on the appropriate analytic harmonic
superfields, some analogs of $q^{+a}\,$,
i.e. whether their defining constraints are also a disguised $\tau$ frame
form of the harmonic analyticity conditions.

\section*{Acknowledgements}
The work of E.I. was supported in part by the RFBR grant 06-02-16684,
the RFBR-DFG grant 06-02-04012-a, the grant DFG, project 436 RUS 113/669/0-3,
the grant INTAS 05-7928
and a grant of Heisenberg-Landau program. He thanks Laboratoire
de Physique, UMR5672 of CNRS and ENS Lyon, for
the warm hospitality extended to him during the course of this work.

\end{document}